\documentclass[twocolumn]{jpsj3}
\usepackage{txfonts}
\usepackage[dvipdfmx]{graphicx}
\usepackage{dcolumn}
\usepackage{bm}
\usepackage{color}
\usepackage{ulem}
\usepackage{widetext}

\title{Spontaneous symmetry breaking of domain walls in phase-competing regions}

\author{Hiroaki Ishizuka$^1$\thanks{ishizuka@appi.t.u-tokyo.ac.jp}, Yasusada Yamada$^2$, 
and Naoto Nagaosa$^{1,3}$}
\inst{
$^1$ Department of Applied Physics, The University of Tokyo, Bunkyo, Tokyo, 113-8656, JAPAN\\
$^2$ Faculty of Engineering Science, Osaka University, Toyonaka, Osaka, 560-8531, JAPAN\\
$^3$RIKEN Center for Emergent Matter Science (CEMS), Wako, Saitama, 351-0198, JAPAN
} 

\abst{
  We study the nature of domain walls in an ordered phase in the phase-competing region of two Ising-type order parameters. Considering a two-component $\phi^4$ theory, we show that the domain wall of the ground-state (primary) order parameter shows a second-order phase transition associated with the secondary order parameter of the competing phase; the effective theory of the phase transition is given by the Landau theory of Ising-type phase transition. We find that the phase boundary of this phase transition is different from the spinodal line of the competing order. Experimentally, the phase transition is detected by the divergence of the susceptibility corresponding to the secondary order when the temperature is quenched to introduce the domain walls.
}

\begin{document}
\maketitle

{\it Introduction} --- Spontaneous symmetry breaking is one of the basic principles in physics. It is described by the order parameters which becomes nonzero in the symmetry broken phase in the thermodynamic limit, i.e., with infinitely large number of particles. In condensed matter physics, the electron correlation often leads to variety of symmetry breaking. For example, in transition metal compounds, the existence of various degrees of freedoms such as spins, orbitals, coordinating anions (lattice degrees of freedom), etc., opens up the possibility for a variety of long-range orderings and phase transitions. Owing to the existence of many different candidate orders, phase coexistence/competition between the different states is often found in these materials~\cite{Dagotto2005}. This often bring about rich physics such as colossal magnetoresistance~\cite{Ramirez1997,Tokura2006} and multiferroics~\cite{Fiebig2005,Tokura2014}. The coexistence of multiple order parameters also play an interesting role in the domain walls. For instance, when two order parameters $\phi_1$ and $\phi_2$ exists in the ground state [suppose the critical temperature for $\phi_1$ ($T_1$) is higher than that of $\phi_2$ ($T_2$)], it was discussed that the domain walls of $\phi_1$ show nonzero $\phi_2$ even above $T_2$ (Refs.~\citen{Houchmandzadeh1991,Daraktchiev2008}).

On the other hand, a similar physics may appear even in the case of phase competition. In SrTiO$_3$, ferroelectric transition in the domain walls was theoretically proposed in the anti-phase boundary, which is potentially relevant to the unusual ferroelectric properties in the low temperature of this material~\cite{Tagantsev2001,Morozovska2012}. The possibility of charge ordering in the phase-competing region between the antiferromagnetic and ferromagnetic phases of La$_{1-x}$Sr$_x$MnO$_3$ was also studied, which could be viewed as a polaron ordering in between the two Jahn-Teller distorted domains~\cite{Yamada1996}. These observations potentially suggest that the phase competition also brings about interesting properties of the domain walls different from the conventional understanding.

\begin{figure}
  \includegraphics[width=\linewidth]{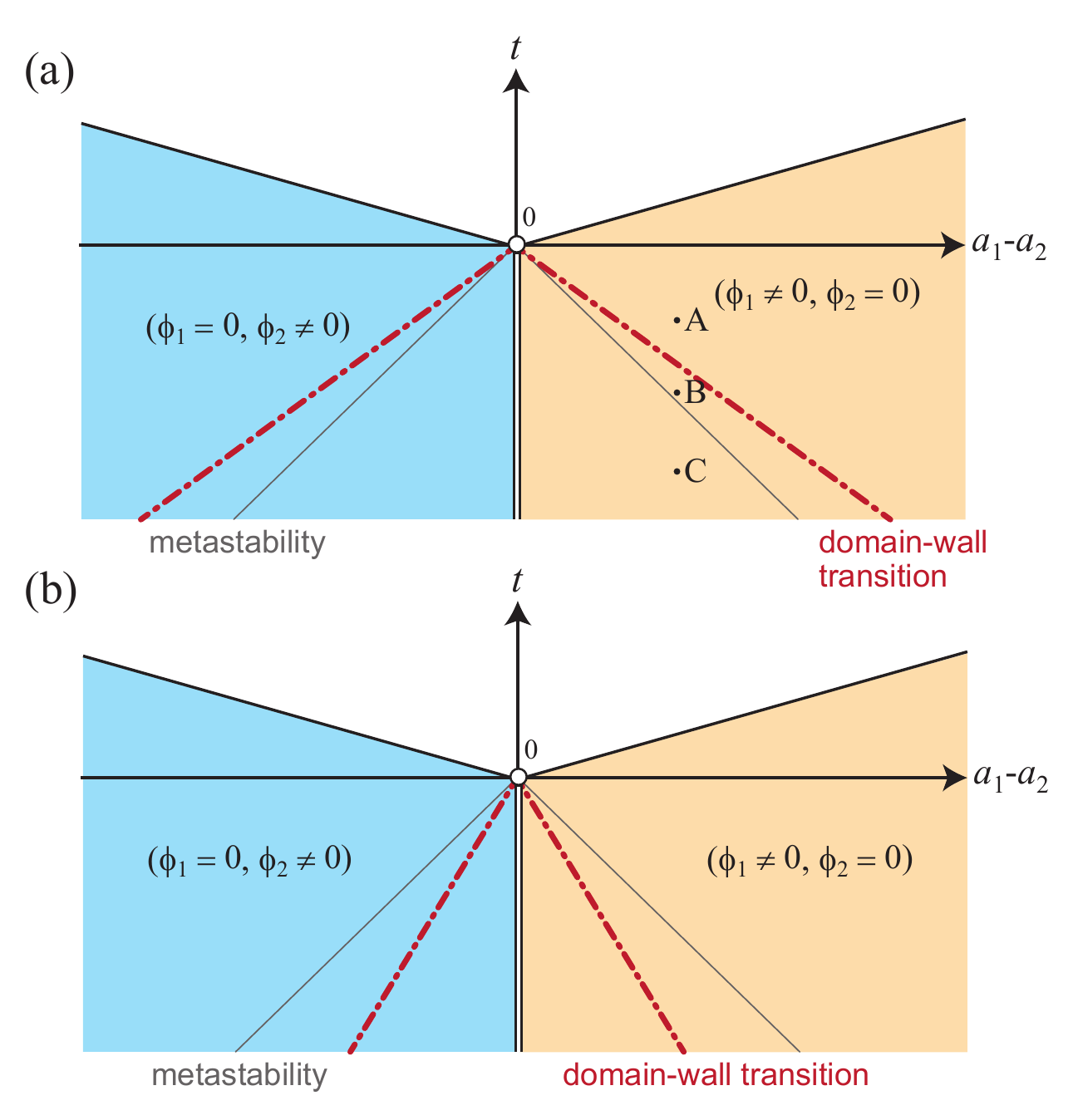}
  \caption{Phase diagram of the $\phi^4$ model in Eq.~\eqref{eq:model}. Phase diagrams in $(a_1-a_2)-t$ space ($a_1,a_2:$ the coefficient of the quadratic terms in eq.(2) of the main text, and $t=-(a_1+a_2)/2$) when the competition is strong enough such that the two orders are separated by the first-order phase transition (the double line along the $t$ axis). The black thick lines show the onset of primary order with second-order phase transition (at the mean-field level), while the thin black lines the limit of metastability of the secondary order, i.e., spinodal line. The red dashed-dotted lines indicate the second-order phase transition due to the onset of the secondary order in the domain walls. The dots A, B, and C, in (a) shows the point that corresponds to Figs.~\ref{fig:Vpp}(a), \ref{fig:Vpp}(b), and \ref{fig:Vpp}(c), respectively. There are two cases, i.e., the domain-wall phase transition occurs without the metastability of the secondary order as shown in (a), or with the metastability as in (b). See the main text on the discussion on Fig.~\ref{fig:phasediagram2}.}\label{fig:phasediagram}
\end{figure}

In an attempt to understand the interplay of phase competition and domain walls, in this Letter, we study two-component real $\phi^4$ theory~\cite{Houchmandzadeh1991}. We find that the domain walls of the primary (ground-state) order parameter show a continuous phase transition associated to the secondary order parameter, which is the competing order parameter of the primary one. This is summarized in Fig.~\ref{fig:phasediagram}. The black lines in the phase diagram correspond to the phase boundary; the two shaded regions are ordered phases and the white region at $t>0$ is the disordered state. The thin lines are the spinodal lines for the metastable state. The competing order become a metastable state inside the cone. On the other hand, the red dashed lines represent the boundary lines of the second-order phase transitions (onset of the secondary order) in the domain walls. Interestingly, the boundary lines are independent of the spinodal line which is associated with the existence of the metastable state. It could be above or below the spinodal lines as shown in Fig.~\ref{fig:phasediagram}(a) and (b), respectively. This result implies that the phase transition in the domain walls are a phenomena that has different physics independent from the metastability.

\begin{figure}
  \includegraphics[width=\linewidth]{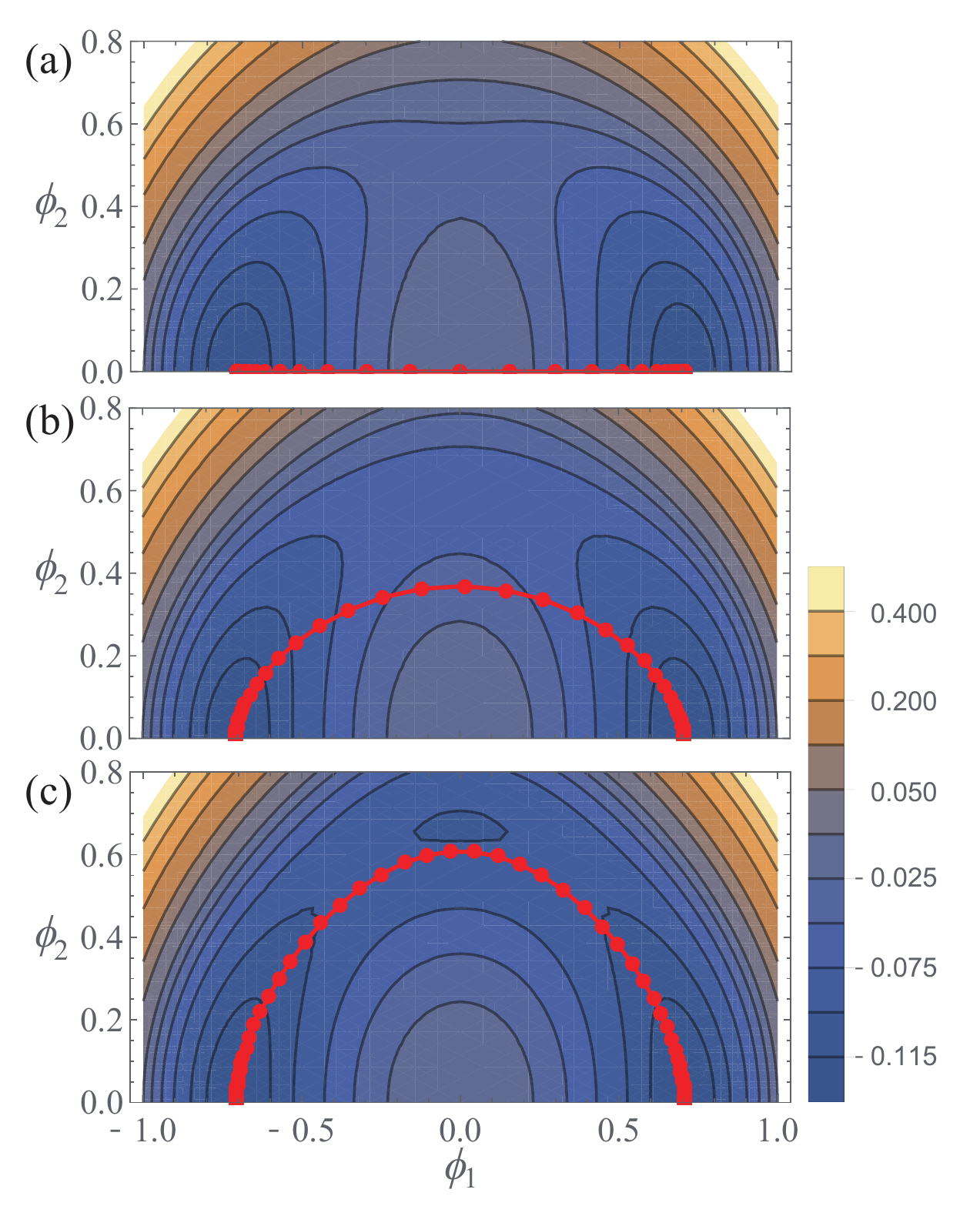}
  \caption{
	 The contour map of $V(\phi_1,\phi_2)$ in the $(\phi_1,\phi_2)$-plane. The points $(\pm\phi_0,0)$ correspond to the ground states of the system. The lines connecting the stable states express the trajectories of the representative points of the system as the system traverses between the two stable domains: (a) $\alpha\equiv a_2/a_1=0.5$, (b) $\alpha=0.7$, and (c) $\alpha=0.9$. (a), (b), and (c) respectively corresponds to A, B, and C in Fig.~\ref{fig:phasediagram}(a).
  }\label{fig:Vpp}
\end{figure}

{\it Model} ---  To study the phase transition in the boundary, we consider a two component $\phi^4$ theory in 1$d$:
\begin{align}
  F=&\int dx\,\left\{\frac{c_1}2(\partial_x\phi_1)^2+\frac{c_2}2(\partial_x\phi_2)^2+V(\phi_1,\phi_2)\right\},\label{eq:model}
\end{align}
where
\begin{align}
  V(\phi_1,\phi_2)=&-\frac{a_1}2\phi_1^2-\frac{a_2}2\phi_2^2+\frac{b_{11}}4(\phi_1^4+\phi_2^4)+\frac{b_{12}}2\phi_1^2\phi_2^2.\label{eq:V(p,p)}
\end{align}
Here, $(\phi_1,\phi_2)=(\phi_1(x),\phi_2(x)\,)\in\mathbb R^2$ is the two component real field and $a_i$, $b_{ij}$, $c_i \in \mathbb R$ ($i,j=1,2$) are real coefficients. The landscape of $V(\phi_1,\phi_2)$ is illustrated in Fig.~\ref{fig:Vpp} in the form of the equi-energy contour map. Physically, the two fields $\phi_1$ and $\phi_2$ correspond to different order parameters, for instance, magnetization and polarization. In this paper, we assume $b_{12}>b_{11}>0$, and $c_i>0$. The ground state phase diagram for these choices of parameters are shown in Fig.~\ref{fig:phasediagram}(a); we assumed $t=-(a_1+a_2)/2$ as the temperature. As shown in the figure, when $a_1>a_2$ and $a_1>0$, the ground state reads $(\phi_1,\phi_2)=(\pm\phi_0,0)$ where $\phi_0=\sqrt{a_1/b_{11}}$ (yellow region). On the other hand, $\phi_1=0$ and $\phi_2\ne0$ phase become the ground state when $a_2>a_1>0$. These two phases are separated by a first-order phase transition (double line); the phase transition between the ordered and the disordered phases [white region in Fig.~\ref{fig:phasediagram}(a)] are of second order. In the rest of the paper, we focus on the $a_1>a_2>0$ case (yellow region) and consider the domain wall in between $(\phi_0,0)$ and $(-\phi_0,0)$ domains.

The countour map of $V(\phi_1,\phi_2)$ in the $(\phi_1,\phi_2)$ plane is given in Fig.~\ref{fig:Vpp}, for each point $A$, $B$, and $C$, in Fig.~\ref{fig:phasediagram}. The red lines indicate the trajectory of the domain wall solution $\vec\phi_{\rm DW}\equiv(\phi_1,\phi_2)$ in the domain wall region, and the dots on them specify the interval in real space coordinate $x$. Figure~\ref{fig:Vpp}(a) corresponds to the case where the secondary order $\phi_2$ does not appear. In Figs.~\ref{fig:Vpp}(b) and \ref{fig:Vpp}(c), on the other hand, the secondary order parameter appears independent of the absence (b) or presence (c) of the metastable state.

{\it Domain wall solution for $\phi_2\ll\phi_1$} --- We here focus on the domain walls between the two domains of the yellow region in Fig.~\ref{fig:phasediagram}(a), and focus on the case $\phi_2(x)\ll\phi_0$. Assuming $\phi_2(x)=0$ for arbitrary $x$, the problem of solving the domain-wall solution reduces to that of the single component $\phi^4$ model; the exact solution reads\cite{Rajaraman1982}
\begin{align}
  \phi_{\rm DW}^0(x)\equiv\pm\phi_0\tanh\left(\frac{x}{2\xi}\right),
\end{align}
where $\xi=\sqrt{c_1/(2a_1)}$. We here assumed that the domain wall is located at $x=0$, since this assumption does not reduce the generality. The local instability of the $\phi_2=0$ solution is examined by expanding the free energy in Eq.~\eqref{eq:model} with respect to $\phi_2(x)$, assuming $\phi_1(x)=\phi_{\rm DW}^0(x)$. The expansion reads
\begin{align}
  F=&\int dx \left[c_2\phi_2\left\{ -\frac12\frac{d^2}{dx^2}-\frac{a_2}{2c_2}+\frac{b_{12}}{2c_2}(\phi_{\rm DW}^0)^2\right\}\phi_2+{\cal O}(\phi_2^4)\right]
\nonumber\\
  &+\quad {\rm const.}\label{eq:F2}
\end{align}
We note that $\delta\phi_1\equiv\phi_1(x)-\phi_{\rm DW}^0(x)$ is implicitly included in the quartic correction term $O(\phi_2^4)$, as the correction to the free energy by $\delta\phi_1(x)$ does not appear in the quadratic order of $\phi_2$.

We further transform the basis for $\phi_2$ in Eq.~\eqref{eq:F2} using the solutions for a Sturm-Liouville equation
\begin{align}
  \left\{ -\frac12\frac{d^2}{dx^2}-\frac{v}{\cosh^2(x/2\xi)}\right\}f_n(x)=\varepsilon_n f_n(x),\label{eq:schroedinger}
\end{align}
where $v=b_{12}(\phi_0)^2/(2c_2)$, $\varepsilon_n$ is the eigenvalue for the $n$th eigenstate ($\varepsilon_n\ge\varepsilon_m$ for $n>m$), and $\phi_n$ is the eigenfunction. For a sufficiently small $n$ the eigenvalues of Eq.~\eqref{eq:schroedinger} is known to be $\varepsilon_n<0$. More precisely, it is known that there are $\lceil s\rceil$ solutions with negative $\varepsilon_n$ where $s(s+1)=2v\xi$ and $\lceil\cdots\rceil$ is the smallest integer greater or equal to $\cdots$ (Ref.~\citen{Landau1958}). Intuitively, this is seen from the fact that Eq.~\eqref{eq:schroedinger} is equivalent to a 1$d$ Schr\"odinger equation with binding potential. The eigenvalues of Eq.~\eqref{eq:schroedinger} for $n<s$ are~\cite{Landau1958,Houchmandzadeh1991}:
\begin{align}
  \varepsilon_n = -\frac12\left(\frac1{2\xi}\right)^2(s-n)^2.\label{eq:evals}
\end{align}

\begin{figure}
  \includegraphics[width=\linewidth]{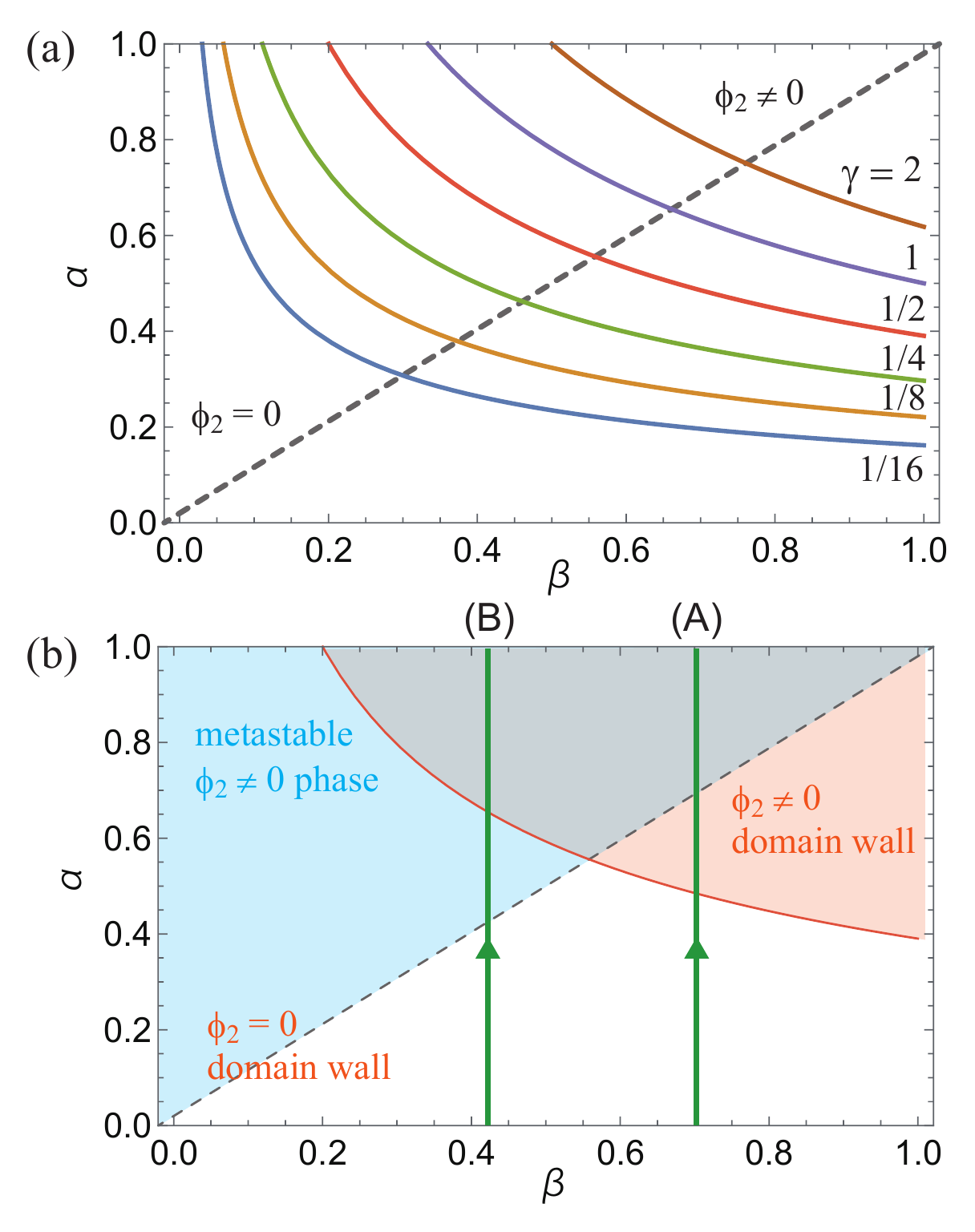}
  \caption{Domain-wall phase diagram of the $\phi^4$ model in Eq.~\eqref{eq:model} in the $\alpha-\beta$ space ($\alpha=a_2/a_1$, $\beta= b_{11}/b_{12}$ in eq.(2)). (a) Each line shows the phase boundary for different $\gamma=c_2/c_1$. Right-top region corresponds to $\vec\phi_{\rm DW}$ with finite $\phi_2$, while left-bottom region to $\phi_2=0$. The dotted line indicates the spinodal line characterizing the metastability of $\phi_2$ order; left-top region corresponds to the case in which the metastable state exists, while right-bottom to the case with saddle points. The $\phi_2\ne0$ region covers all of the phase diagram when $\gamma\to0$, while it shrinks to $(\alpha,\beta)=(1,1)$ for $\gamma\to\infty$. (b) Schematic figure illustrating the relation between the phase diagram in (a) and those in Fig.~\ref{fig:phasediagram}. The green lines show trajectories with decreasing temperature: lines (A) and (B) respectively correspond to the phase diagrams (a) and (b) in Fig.~\ref{fig:phasediagram}.}\label{fig:phasediagram2}
\end{figure}

Using this basis, the Free energy in Eq.~\eqref{eq:F2} reads
\begin{align}
  F &= \sum_{n=0} c_2\left\{\varepsilon_n+v-\frac{a_2}{2c_2}\right\}\psi_n^2
+{\cal O}(\psi_n^4)+{\rm const.},\label{eq:F2-2}
\end{align}
where $\phi_2(x)=\sum_n \psi_n f_n(x)$ and $\varepsilon_n$ for $n<s$ is given in Eq.~\eqref{eq:evals}. Therefore, the $\phi_2=0$ solution become unstable when $\varepsilon_0<\frac{a_2}{2c_2}-v$; the condition reads
\begin{align}
  \alpha>\chi\equiv\frac\gamma4\left(\sqrt{1+\frac8{\gamma\beta}}-1\right).\label{eq:pdboundary}
\end{align}
Here, $\alpha=a_2/a_1$, $\beta=b_{11}/b_{12}$, and $\gamma=c_2/c_1$. The phase diagram of the domain-wall solution is shown in Fig.~\ref{fig:phasediagram2}(a). Here, due to the condition for $a_i$ and $b_{ij}$, $\alpha,\beta\in(0,1)$. Each line in the figure shows the phase boundary for different $\gamma$; the phase boundary approaches $(1,1)$ as $\gamma\to\infty$ while all region will be covered by $\phi_2\ne0$ phase for $\gamma\to0$ (The phase boundary approaches $\alpha=0$ and $\beta=0$ lines as $\gamma\to0$.). In our model, $c_2$ is the stiffness of $\phi_2$ while the width of the domain wall is $2\xi\propto\sqrt{c_1}$. Therefore, intuitively, the suppression of $\phi_2\ne0$ phase with increasing $\gamma$ reflects the fact that it costs the energy to deform the $\phi_2$ field when $\gamma$ is increased. Instead, when $\gamma$ is decreased, the total domain wall energy can be reduced by favoring $V(\phi_1,\phi_2)$ relative to the gradient energy, thus the trajectory tends to take on a `detour' around the local maximum at $(\phi_1,\phi_2)=(0,0)$ (See Fig.~\ref{fig:Vpp}). We also note that this phase boundary is different from the spinodal line for the metastable state $(\phi_1,\phi_2)=(0,\pm\sqrt{a_2/b_{11}})$; this state is metastable when $\alpha>\beta$. As shown by the dotted line in Fig.~\ref{fig:phasediagram2}(a), the spinodal line has no relation to the domain-wall phase boundary.

Substituting $\alpha$ in Eq.~\eqref{eq:pdboundary} by $t$, the $\phi_2=0$ domain wall become unstable when
\begin{align}
  t<-\frac{1+\chi}{2(1-\chi)}|a_1-a_2|,\label{eq:boundaryTc}
\end{align}
where $\chi$ is given in Eq.~\eqref{eq:pdboundary}. Similarly, the spinodal line is given by the same equation, by replacing $\chi\to\beta.$ In Fig.~\ref{fig:phasediagram}(a), the red dashed line on the right hand side of the figure shows the phase boundary for the domain wall, below which a nonzero $\phi_2$ appears at the domain walls of $\phi_1$. Eq.~\eqref{eq:boundaryTc}, shows that the critical temperature for the domain-wall phase transition approaches that of the bulk critical temperature as $(a_1-a_2)\to0$.

\begin{figure}
  \centering
  \includegraphics[width=\linewidth]{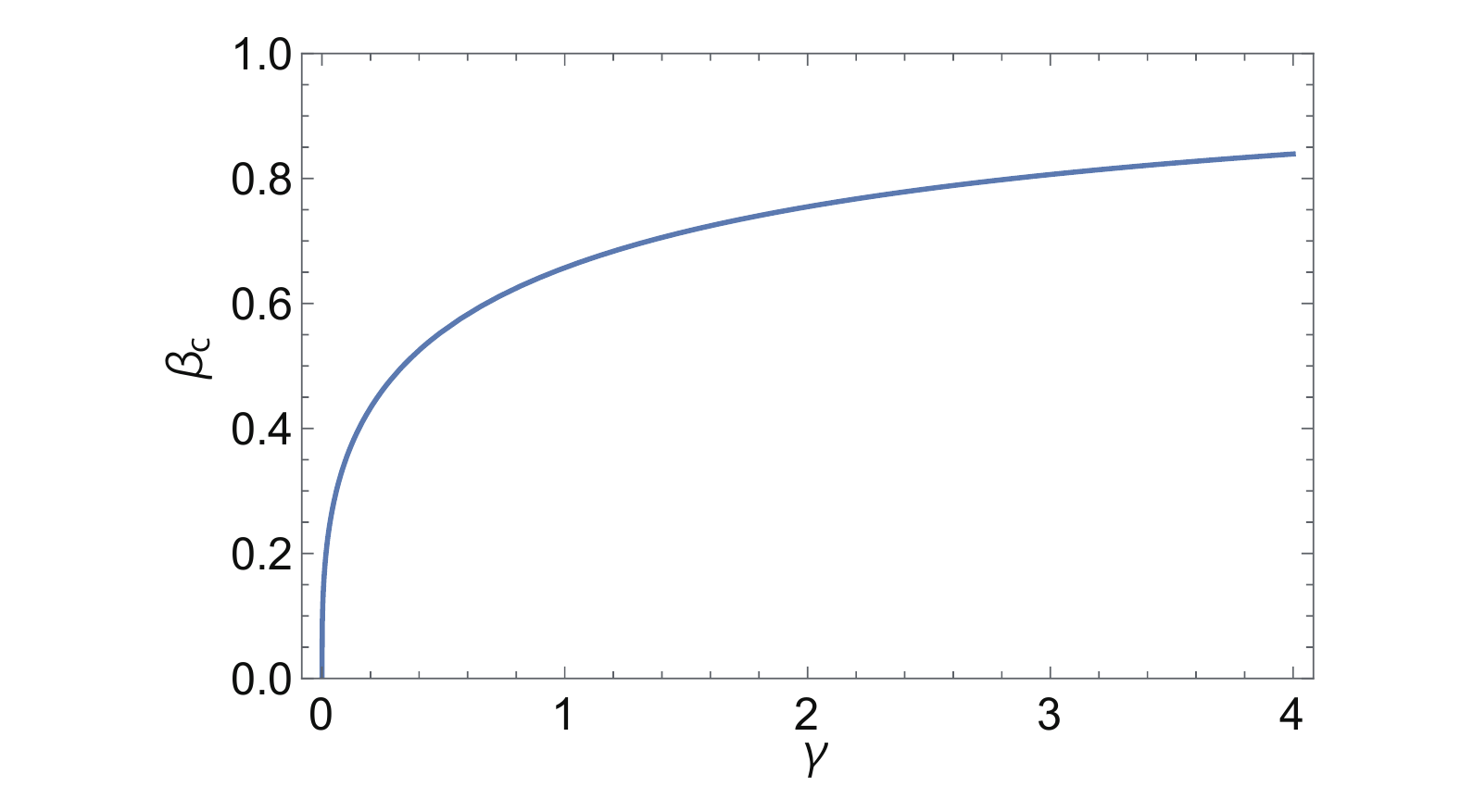}
  \caption{The $\gamma$ dependence of $\beta_c$ in Eq.~\eqref{eq:betaC}. The function goes to $\beta_c\to0$ as $\gamma\to0$, while it approaches $\beta_c\to1$ in the $\gamma\to\infty$ limit.}\label{fig:betaC}
\end{figure}

For a given $a_1-a_2$, the spinodal line can be at a lower or higher temperature than the domain-wall critical point. This is governed by the value of $\beta$; the domain-wall critical point is at a higher temperature when $\beta>\beta_c$ while the spinodal line is at a higher temperature for $\beta<\beta_c$. The phase diagram for $\beta>\beta_c$ is plotted in Fig.~\ref{fig:phasediagram}(a). On the other hand, the case for $\beta<\beta_c$ is shown in Fig.~\ref{fig:phasediagram}(b). The critical value $\beta_c$ reads
\begin{align}
  \beta_c&=\frac16\left(\frac{\gamma^2}{\omega(\gamma)}+\omega(\gamma)-\gamma\right)\label{eq:betaC}
\end{align}
where
\begin{align}
  \omega(\gamma)&=\left(54\gamma-\gamma^3+6\sqrt{(9\gamma)^2-3\gamma^4}\right)^{\frac13}\equiv \lambda^{\frac13}.
\end{align}
Here, $\lambda^{\frac13}$ is the triple root of $\lambda$ with the smallest non-negative argument. The $\gamma$ dependence of $\beta_c$ is plotted in Fig.~\ref{fig:betaC}. As shown in the figure, $\beta_c$ in Eq.~\eqref{eq:betaC} is a monotonic function of $\gamma$ where $\beta_c\to0$ when $\gamma\to0$ and $\beta_c\to1$ when $\gamma\to\infty$. These features are diagrammatically seen from the abstract phase diagram in Fig.~\ref{fig:phasediagram2}(b). In Fig.~\ref{fig:phasediagram2}(b), we show the phase boundary for $\gamma=1/2$ and the spinodal line. Assuming $b_{11}$ and $b_{12}$ are fixed (hence, $\beta$ is fixed), reducing $t$ corresponds to increasing $\alpha$ in the phase diagrams [green lines in Fig.~\ref{fig:phasediagram2}(b)]. In this figure, $\beta_c$ corresponds to the $\beta$ where the spinodal line crosses the domain-wall phase boundary. Hence, the domain-wall critical temperature is higher when $\beta>\beta_c$ and vice versa, and $\beta_c$ changes from 0 to 1 with increasing $\gamma$.

{\it Phase transition in the domain wall} --- Another important aspect of the two different kinds of domain walls is the degeneracy; $\phi_2=0$ phase has a unique domain wall symmetric under $\phi_2\to-\phi_2$ while the $\phi_2\ne0$ domain wall has two degenerate domain walls which are connected by $\phi_2\to-\phi_2$ operation. This nature of the domain walls resemble that of the bulk property of the ferromagnetic Ising model, which the paramagnetic and ferromagnetic phases are separated by a second-order phase transition. Indeed, by assuming $\psi_n=0$ for $n\ge 1$, the low energy model for the domain wall reads
\begin{align}
  F &\sim F_0\equiv c_2\left\{\varepsilon_0-\frac{a_2}{2c_2}+v\right\}\psi_0^2+\frac{u}4\psi_{0}^4,
\end{align}
where $u$ is the coefficient for the quartic order term for $\psi_0$. This free energy is identical to the Landau theory of second-order phase transition. Therefore, we expect the susceptibility for $\psi_0$ to diverge at the phase boundary in the same manner as in the Landau theory. This is consistent with the initial assumption that $\phi_2$ is sufficiently small compared to $\phi_1$ close to the boundary of the domain-wall phase diagram. We also note that, a general property of the Sturm-Liouville equation ensures $f_0(x)$ is non-negative (More precisely, it is nodeless.). Hence, the susceptibility of $\psi_0$ is proportional to the susceptibility of $\phi_2$ when the system is close to the domain-wall phase boundary. Therefore, the phase transition in the domain walls should be experimentally detected by the diverging susceptibility of the associated order parameter $\phi_2$.

We also note that there could be another phase transition associated with $\psi_1$ below the critical temperature for $\psi_0$. If we take into account of the contribution from $\psi_1$ in addition to $F_0$, the general form of the free energy reads:
\begin{align}
  F \sim F_0+ c_2\left\{\varepsilon_1-\frac{a_2}{2c_2}+\nu\right\}\psi_1^2+\frac{u'}4\psi_{1}^4+\frac32u''\psi_0^2\psi_1^2.
\end{align}
Here, $u'$ and $u''$ are real coefficients for the quartic order terms of $\psi_1$. The free energy above does not have a term that couples to $\psi_1$ linearly. Therefore, a phase transition associated with $\psi_1$ is also possible. This is a consequence of the symmetry of the Sturm-Liouville equation in Eq.~\eqref{eq:schroedinger}. As the equation is symmetric with respect to the operation $x\to-x$, the eigenfunctions have either even or odd parity with respect to the operation. Furthermore, the general theory of Sturm-Liouville equation ensures the existence of one node for $\phi_1(x)$. These two facts manifest $\phi_1(x)$ is a odd function of $x$ with the node at $x=0$. Therefore, the symmetry prohibits the terms of form $f(\psi_0)\psi_1$, where $f(x)$ is an arbitrary real function. From the viewpoint of the theory of phase transition, the phase transition associated with $\psi_1$ breaks the mirror symmetry $x\to-x$. For $\psi_n$ ($n\ge2$), however, the phase transition similar to $\psi_1$ does not exists in general. For these $\psi_n$, in general, there exists a term like $\psi_0^3\psi_n$ for an even $n$; similarly, $\psi_1^3\psi_n$ exists for an odd $n$. Therefore, no phase transitions associated with these parameters occur in general.

{\it Effect of meandering} --- So far, we have focused on the 1$d$ theory. The same argument, however, holds for higher dimensions assuming $\vec\phi_{\rm DW}$ is uniform along the other directions. This potentially provides an experimental setup to study a strongly correlated systems in the confined dimension. In experiment, however, it is difficult to prepare a flat domain wall. For instance, the domain walls generated as the phase boundary of domains in thermal quenching experiments are meandered, and is usually difficult to shape them flat. Theoretically, it is expected that the effect of such meandering is effectively taken into account as the random temperature. However, it has been discussed that such a disorder does not change the critical behavior except for some logarithmic corrections~\cite{Shalaev1984,Shankar1987,Ludwig1988,Ludwig1990,Wang1990a,Shalaev1994}. Therefore, we expect such meandering does not affect the phase transitions, at least qualitatively. For $\psi_1$, however, the phase transition becomes a crossover as the transition is associated with the symmetry $x\to-x$, i.e., the mirror operation about the plane along the domain wall, which is absent in the presence of meandering.

{\it Conclusion} --- To conclude, in this Letter, we studied the phase diagram of the domain wall for the two component real $\phi^4$ theory for the case of $a_1>a_2>0$, $b_{12}>b_{11}>0$ and $c_i>0$. We find that in the domain wall region between the two spatially uniform bulk phases $(\phi_0,0)$ and $(-\phi_0,0)$, there exists a {\it second-order} phase transition associated with the onset of the secondary order parameter, $\phi_2$. This phase transition is independent of the presence/absence of the metastable state. This is manifested by the fact that our treatment on domain wall system is based on an effective Ginzburg-Landau theory which is identical to that of the 2d Ising spin system. Our analysis shows that the critical temperature for the domain-wall phase transition approaches that of the bulk critical temperature as the system approaches the phase boundary between the two ground states. Therefore, the boundary phase transition is expected to be realized in the vicinity of the phase boundary. Experimentally, one needs to introduce the domain walls of the primary order parameter by the quenching across the transition temperature of the bulk. Then the critical phenomenon with $(d-1)$-dimension including the divergence of the susceptibility of the secondary order is realized.

\begin{acknowledgment}
The authors thank T. Kagawa for discussions. This work was supported by JSPS KAKENHI (Grant Nos. JP16H06717 and JP26103006), ImPACT Program of Council for Science, Technology and Innovation (Cabinet office, Government of Japan), and CREST, JST (Grant No. JPMJCR16F1).
\end{acknowledgment}


\begin{thebibliography}{99}
\bibitem{Dagotto2005}            E. Dagotto, Science {\bf 309}, 257 (2005).
\bibitem{Ramirez1997}            A. P. Ramirez, J. Phys.: Condens. Matter {\bf 9}, 8171 (1997).
\bibitem{Tokura2006}             Y. Tokura, Rep. Prog. Phys. {\bf 69}, 797 (2006).
\bibitem{Fiebig2005}             M. Fiebig, J. Phys. D: Appl. Phys. {\bf 38}, R123 (2005).
\bibitem{Tokura2014}             Y. Tokura, S. Seki, and N. Nagaosa, Rep. Prog. Phys. {\bf 77}, 076501 (2014).
\bibitem{Houchmandzadeh1991}     B. Houchmandzadeh, J. Lahzerowicz, and E. Salje, J. Phys.: Condens. Matter {\bf 3}, 5163 (1991).
\bibitem{Daraktchiev2008}        M. Daraktchiev, G. Catalan, and J. F. Scott, Ferroelectrics {\bf 375}, 122 (2008).
\bibitem{Tagantsev2001}          A. K. Tagantsev, E. Courtens, and L. Arzel, Phys. Rev. B {\bf 64}, 224107 (2001). 
\bibitem{Morozovska2012}         A. N. Morozovska, E. A. Eliseev, M. D. Glinchuk, L.-Q. Chen, and V. Gopalan, Phys. Rev. B 85, 094107 (2012).
\bibitem{Yamada1996}             Y. Yamada, O. Hino, S. Nohdo, R. Kanao, T. Inami, and S. Katano, Phys. Rev. Lett. {\bf 77}, 487 (1996).
\bibitem{Rajaraman1982}          R. Rajaraman, {\it Solitons and Instantons}, (Elsevier, Amsterdam, 1982) 1st ed.
\bibitem{Landau1958}             L. D. Landau and E. M. Lifshitz, {\it Quantum Mechanics}, {\it Course of Theoretical Physics}, Vol. 3, (Butterworth-Heinemann, Oxford University Press, 1958) 3rd ed.
\bibitem{Shalaev1984}            B. N. Shalaev, Sov. Phys. Solid State {\bf 26}, 1811 (1984).
\bibitem{Shankar1987}            R. Shankar, Phys. Rev. Lett. {\bf 58}, 2466 (1987).
\bibitem{Ludwig1988}             A. W. W. Ludwig, Phys. Rev. Lett. {\bf 61}, 2388 (1988).
\bibitem{Ludwig1990}             A. W. W. Ludwig, Nucl. Phys. B {\bf 330}, 639 (1990).
\bibitem{Wang1990a}              J.-S. Wang, W. Selke, V. S. Dotsenko, and V. B. Andreichenko, Physica A {\bf 164}, 221 (1990); Europhys. Lett. {\bf 11}, 301 (1990); Nucl. Phys. B {\bf 344}, 531 (1990).
\bibitem{Shalaev1994}            B. N. Shalaev, Phys. Rev. {\bf 237}, 129 (1994).
\end{thebibliography}
\end{document}